\documentclass[pra,twocolumn,showpacs,amsmath,amssymb]{revtex4-1}

\usepackage{lipsum}

\usepackage{amsbsy}
\usepackage{amsfonts}
\usepackage{amssymb}
\usepackage{amsmath}
\usepackage{color}
\usepackage{graphicx}
\begin{document}


\title{A Matrix-Product-Operator Approach to the Nonequilibrium Steady State of Driven-Dissipative Quantum Arrays}

\author{Eduardo Mascarenhas}
\affiliation{Institute of Theoretical Physics, Ecole Polytechnique F\'{e}d\'{e}rale de Lausanne EPFL, CH-1015 Lausanne, Switzerland}

\author{Hugo Flayac}
\affiliation{Institute of Theoretical Physics, Ecole Polytechnique F\'{e}d\'{e}rale de Lausanne EPFL, CH-1015 Lausanne, Switzerland}

\author{Vincenzo Savona}
\affiliation{Institute of Theoretical Physics, Ecole Polytechnique F\'{e}d\'{e}rale de Lausanne EPFL, CH-1015 Lausanne, Switzerland}

\begin{abstract}
We develop a numerical procedure to efficiently model the nonequilibrium steady state of one-dimensional arrays of open quantum systems, based on a matrix-product operator ansatz for the density matrix. The procedure searches for the null eigenvalue of the Liouvillian superoperator by sweeping along the system while carrying out a partial diagonalization of the single-site stationary problem. It bears full analogy to the density-matrix renormalization group approach to the ground state of isolated systems, and its numerical complexity scales as a power law with the bond dimension. The method brings considerable advantage when compared to the integration of the time-dependent problem via Trotter decomposition, as it can address arbitrarily long-ranged couplings. Additionally, it ensures numerical stability in the case of weakly dissipative systems thanks to a slow tuning of the dissipation rates along the sweeps. We have tested the method on a driven-dissipative spin chain, under various assumptions for the Hamiltonian, drive, and dissipation parameters, and compared the results to those obtained both by Trotter dynamics and Monte-Carlo wave function. {Accurate and numerically stable convergence was always achieved when applying the method to systems with a gapped Liouvillian and a non-degenerate steady-state.}
\end{abstract}


\maketitle

\section{Introduction}

The study of the nonequilibrium dynamics of open many-body quantum systems has gained significant momentum in recent years, thanks to the experimental progress achieved in several areas, including ultracold atoms in optical lattices~\cite{UltracoldRev,UltracoldRev2,UltracoldRev3,UltracoldExp,UltracoldExp2,UltracoldExp3,UltracoldExp4}, trapped ions~\cite{Ions,Ions2,Ions3}, arrays of optical micro-resonators~\cite{Cavities,Cavities2,Cavities3} and superconducting circuits~\cite{SuperC,SuperC2,SuperC3}.

A feature common to all these systems is the coupling to an external environment in the form of coherent or incoherent input and output channels. The time evolution of the system is then governed by an interplay of the Hamiltonian and the driven-dissipative dynamics. For stationary external conditions, this dynamics typically leads to a nonequilibrium steady state (NESS), for which a multitude of novel phenomena are expected, including nonequilibrium quantum phase transitions~\cite{StatesPhases,Prosen1,Prosen3,DyBose,SpinPT} and the possibility of engineering quantum states through tailored dissipation~\cite{QSprep}, in view of advanced quantum information strategies~\cite{Comp}.

The theoretical description and modeling of open quantum systems out of equilibrium represents a major challenge. Indeed, similarly to the ground state of isolated many-body quantum systems, the NESS can be characterized by quantum correlations which -- particularly when approaching criticality -- require for their exact determination a computational effort that scales exponentially with the system size~\cite{QPT,QPT2}. As added difficulties however, the NESS is generally not a pure quantum state, nor can it be directly determined from the Gibbs principle, as in the case of thermal equilibrium.

Generally, an open quantum system is described by a density matrix $\hat{\rho}$, whose dynamics obeys the Von Neumann equation ${\dot{\hat{{\rho}}}}={\mathcal L}\hat{\rho}$ dictated by the Liouvillian superoperator ${\mathcal L}$ (we set $\hbar=1$ here and in what follows)~\cite{QM,QNoise}. Two strategies are then available for the determination of the NESS. First, one can directly integrate the time evolution until stationarity is reached. Second, a solution of the equation ${\mathcal L}\hat \rho=0$ can be directly computed, under the additional condition that $\mathrm{Tr}(\hat{\rho})=1$. Apart from special cases in which analytical solutions can be found~\cite{Prosen4,Prosen5}, both strategies can be handled numerically only for very small systems~\cite{Diode,Sarinha,Vinnie,Fermi,Nissen} -- if an exact solution is sought. Larger systems typically require some level of approximation and, still in recent times, many studies have restricted to mean-field approximations~\cite{Nissen,MeanFiled,Solid,Solid2,JCLaser}, thus neglecting quantum correlations. Only very recently a variational principle for the NESS of open quantum systems has been demonstrated~\cite{Weimer} and applied to 1-D systems~\cite{Mari}, while a spatial decimation method specific to the stationary Von Neumann problem has been proposed~\cite{CornerSpace}.

In this scenario, one-dimensional systems represent a special case in which a very accurate description of the many-body quantum state is made possible thanks to the advent of the Density Matrix Renormalization Group~\cite{DMRGWhite,DMRGSch,DMRGGb} (DMRG) and of the equivalent variational approach based on the Matrix Product State (MPS) ansatz~\cite{MPSVers,MPSSch}. In typical situations, the MPS-DMRG approach allows a surprisingly good account of quantum correlations at finite spatial range, with a computational overhead that scales polynomially with the dimension of the Hilbert space. The MPS approach has been successfully extended to the modelling of the unitary time evolution of a closed quantum system~\cite{MPSVers,MPSSch}. For open quantum systems, an analogous Matrix Product Operator (MPO) ansatz for the density matrix has been proposed and applied to model both thermal equilibrium\cite{MPOVer} and temporal dynamics~\cite{MPOVer,MPOVidal}. In particular, the long time dynamics has been employed to obtain the nonequilibrium steady state (NESS) -- in presence of driving fields and dissipation -- in different settings~\cite{Hart,Barthel,Keeling,Bonnes,Clark,ProsenSim}. There are however several settings in which the MPO dynamical approach to the NESS suffers from limitations. This is the case in presence of slow dissipation rates (compared to the energy scale set by the Hamiltonian), or when dissipation acts on a small part of the system only, as in transport configurations~\cite{Clark,ProsenSim,PhtonTrans}. Some systems may even display an algebraic, rather then exponential dynamics to the NESS~\cite{Barthel,Schft1,Schft2,Schft3,Schft4}. Finally the Trotter decomposition, typically used in these dynamical schemes, suffers from a severe limitation: it is restricted to nearest neighbor couplings. Only recently, numerical approaches have been suggested~\cite{Unify,Pollman,LucaSchft1,LucaSchft2}, that overcome this limitation, but only for modeling the unitary dynamics of isolated systems.

Here, we develop an efficient implementation of the variational principle to directly determine the NESS of Markovian open quantum systems. The method does not rely on the integration of the long-time dynamics, thus lifting all the limitations described above. The variational principle for determining the NESS has recently been proposed~\cite{Weimer,Mari} and implemented within an MPS-DMRG scheme~\cite{Mari}. 
{The approach that we propose relies directly on the search for the zero-eigenvalue of the superoperator $\mathcal{L}$ for the determination of the NESS. This approach has shown full numerical stability {when applied to gapped Liouvillians with a non-degenerate NESS}}. As a test of the method, we simulate a driven-dissipative Ising chain, and compare the results to those obtained by simulating the MPO dynamics~\cite{Ima,Keeling} and with Monte-Carlo Wave Function (MCWF)~\cite{MCWF1,MCWF2,MCWF3,MCWF4}. We then simulate the same system, in presence of longer-range couplings or slow dissipation rates, thus showing its wide range of applicability in the description of driven dissipative systems. {We finally discuss the computational complexity of the approach and compare it to other existing methods~\cite{Mari}.}

\section{The method}

We consider a one-dimensional chain of $N$ coupled quantum systems, each characterized by $d$ possible states, in the presence of external driving fields and Markovian coupling to the external environment. The dynamics is governed by the Lindblad-Von Neumann master equation~\cite{QM,QNoise}
\begin{equation}
\frac{d\hat\rho}{dt}=\mathcal{L}\hat\rho=-i[{\mathcal{\hat H}},\hat\rho]-\frac{1}{2}\sum_i\left[\{\hat K_i^{\dagger}\hat K_i,\hat\rho\}-2\hat K_i\hat\rho \hat K_i^{\dagger}\right]\,,
\end{equation}
where ${\hat K}_i$ are the operators corresponding to the transitions induced by the environment. The NESS solution obeys the equation $\mathcal{L}\hat{\rho}_{\mathrm{NESS}}=0$.

For the purpose of numerical implementation, it is convenient to map the MPO representation onto an equivalent MPS form. We do this by the vectorization procedure, where the density matrix $\hat\rho$ is reshaped into a column vector, here denoted by $|\hat\rho \rangle\rangle$, by concatenating all its columns. To express the Liouvillian superoperator in this representation, we rely on the property $|X\hat\rho Y \rangle\rangle=Y^{T}\otimes X|\hat\rho \rangle\rangle$, where $X$ and $Y$ are matrices. Then, $\mathcal{L}$ takes the form of the matrix defined by~\cite{LM}:
\begin{eqnarray}
\mathcal{L}&=&-i(\mathbb{I}\otimes {\mathcal{\hat H}}-{\mathcal{\hat H}}^{T}\otimes \mathbb{I})\nonumber\\
&+&\frac{1}{2}\sum_i (2\hat K_i^{\ast}\otimes \hat K_i-\mathbb{I}\otimes \hat K_i^{\dagger}\hat K_i-\hat K_i^{T}\hat K_i^{\ast}\otimes \mathbb{I})\,.
\end{eqnarray}

The determination of the NESS can then be reformulated as the variational minimization of the euclidean norm functional
\begin{equation}
||\mathcal{L}|\hat\rho \rangle\rangle||\ge0\,.\label{Func}
\end{equation}
\begin{figure}
\includegraphics[width=1\linewidth]{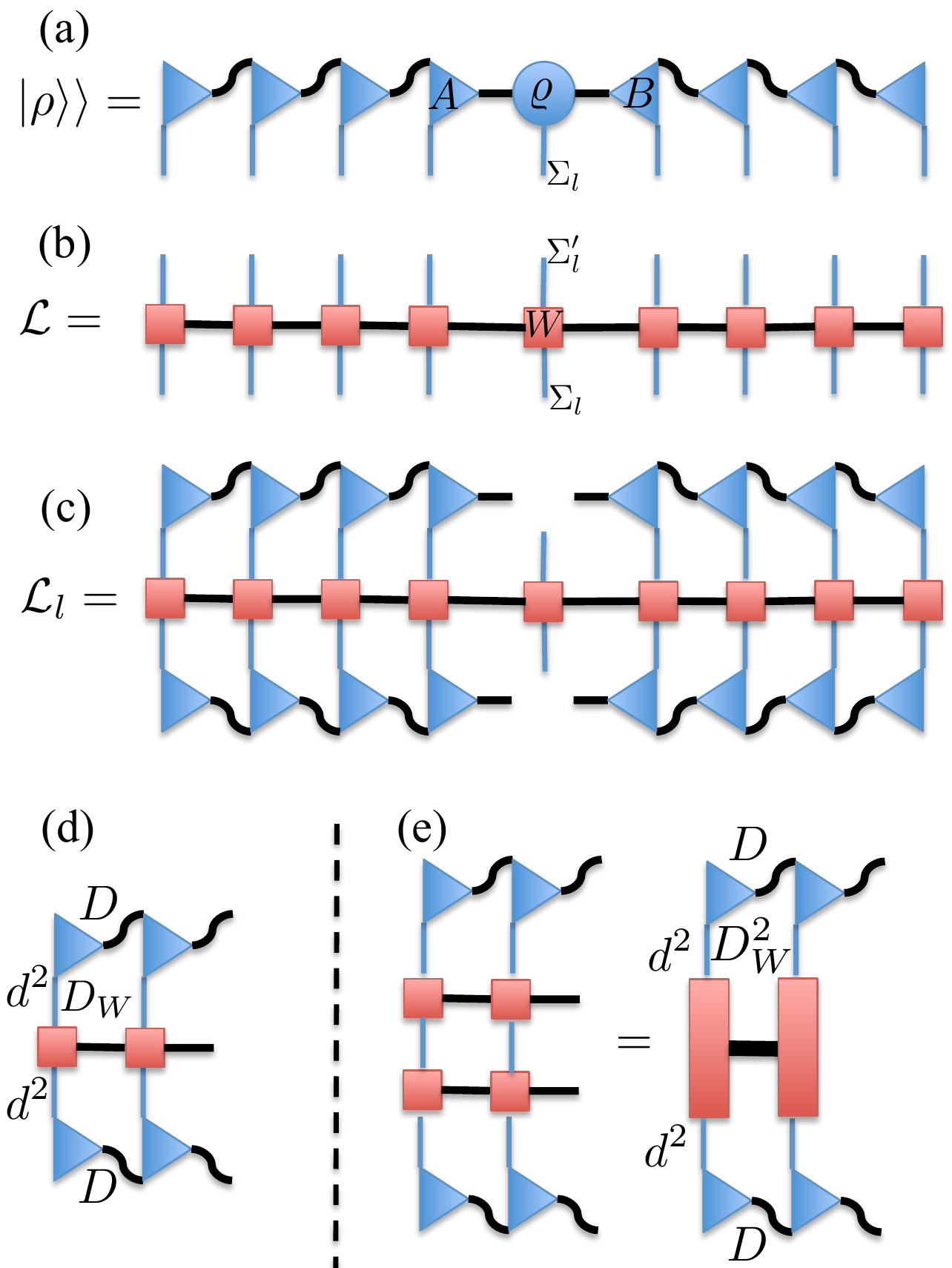}
\caption{Diagrammatic representations of matrix products. (a) Diagram of the vectorized density matrix as an MPS in the mixed canonical form (\ref{RhoMixed}). (b) Diagram of the Liouvillian operator in the MPO representation (\ref{LindbladMPO}). (c) Diagram representing the on-site Liouvillian operator $\mathcal{L}_l$. In all diagrams, triangles pointing right represent the left-normalized matrices $A$, while triangles pointing left denote the right-normalized matrices $B$, both entering the mixed canonical form of the MPS, Eq. (\ref{RhoMixed}). The local representation of the density matrix $\varrho$ is depicted as a circle at site $l$. Thin lines represent physical indices while thick lines denote bond indices. The MPO matrices $W$ in Eq. (\ref{LindbladMPO}) are represented as squares. (d) Diagrammatic scheme illustrating the computational complexity associated with index contractions at one site, in the case where the simple Liouvillian $\mathcal{L}$ is used in the variational approach. (e) Same as (d), in the case where the squared Liouvillian  $\mathcal{L}^{\dagger}\mathcal{L}$ is instead used. In this second case, contraction of the MPO bond indices bears an additional $O(D_W)$ computational cost.}\label{Graph}
\end{figure}

The MPO representation of the density matrix reads
$\hat\rho = \sum_{\boldsymbol{\sigma}\boldsymbol{\sigma}'}\left(\prod_{i=1}^{N} {A}^{\sigma_i\sigma_i'}\right)|\boldsymbol{\sigma}\rangle\langle\boldsymbol{\sigma}'|$,
in which $|\boldsymbol{\sigma}\rangle=|\sigma_1\ldots\sigma_l\ldots \sigma_N\rangle$ are the states of the system, $|\sigma_j\rangle$ is the state of the $j$-th site of the chain and the sets of matrices $\{A\}$ parametrizes the MPO state~\cite{MPSVers,MPSSch}. Through vectorization, we may express the density matrix as an MPS
\begin{equation}
|\hat\rho\rangle\rangle = \sum_{\boldsymbol{\Sigma}}\left(\prod_{i=1}^{N} {A}^{\Sigma_i}\right)|\boldsymbol{\Sigma}\rangle\rangle\,,
\end{equation}
where $|\boldsymbol{\Sigma}\rangle\rangle=||\boldsymbol{\sigma}\rangle\langle\boldsymbol{\sigma}'|\rangle\rangle$, and the indices in the matrix elements of $\hat{\rho}$ have been encoded as $\Sigma_i=(\sigma_i'-1)d+\sigma_i$. Once expressed using a MPS representation, the problem is determined by (\ref{Func}) can be solved using the MPS-DMRG strategy, for which we will refer to the treatment -- and the related notation -- extensively presented in Ref.\onlinecite{MPSSch}. In particular, in order to derive the equation for the on-site problem, it is useful to express the density matrix in a mixed canonical form~\cite{MPSSch}
\begin{equation}
|\hat\rho\rangle\rangle = \sum_{\boldsymbol{\Sigma}}\left(\prod_{i=1}^{l-1} {A}^{\Sigma_i}\right){\varrho}^{\Sigma_l}\left(\prod_{i=l+1}^N {B}^{\Sigma_i}\right)|\boldsymbol{\Sigma}\rangle\rangle\,,\label{RhoMixed}
\end{equation}
where the matrices ${\varrho}^{\Sigma_l}$, associated to the $l$-th site in the MPS ansatz, have been singled out from the MPS expression, and the sets of matrices $\{A\}$ and $\{B\}$, with maximal bond dimension $D$, are left and right normalized, respectively~\cite{MPSVers,MPSSch}. The MPS is depicted in Fig.~\ref{Graph}(a) in the usual diagrammatic representation~\cite{MPSVers,Unify}. The symbol $\varrho$ then denotes a rank-three tensor, and is associated to a local representation of the density matrix at site $l$.

The Liouvillian operator can be represented in an MPO form as
\begin{equation}
\mathcal{L}=\sum_{\boldsymbol{\Sigma},\boldsymbol{\Sigma}'}\prod_{i=1}^N W^{{\Sigma_i}{\Sigma_i}'}|\boldsymbol{\Sigma}\rangle\rangle\langle\langle\boldsymbol{\Sigma}'|\,,\label{LindbladMPO}
\end{equation}
as depicted in Fig.~\ref{Graph}(b). Here, $D_W$ is the bond dimension of the MPO representation of $\mathcal{L}$, i.e. the dimension of the matrices $W$ in (\ref{LindbladMPO}). $D_W$ is defined by the complexity of the system Hamiltonian and dissipative processes and is fixed for a given model~\cite{MPSSch}.

A most natural choice for the variational determination of the NESS, as adopted in Ref.~\onlinecite{Mari}, would be to express (\ref{Func}) as $\langle\langle\hat\rho|\mathcal{L}^{\dagger}\mathcal{L}|\hat\rho\rangle\rangle$. In this way, the problem bears a full analogy to the MPS-DMRG approach to isolated systems, with the hermitian, semi-positive-defined operator $\mathcal{L}^\dagger\mathcal{L}$ playing the role of the Hamiltonian. {However, this choice requires handling the product $\mathcal{L}^{\dagger}\mathcal{L}$ at some level within the algorithm. Let us assume a given MPO representation of $\mathcal{L}$, with bond dimension $D_W$. As depicted in Fig.~\ref{Graph}(d), when computing the quantity $\langle\langle\hat\rho|\mathcal{L}|\hat\rho\rangle\rangle$ the numerical complexity associated to the index contractions on each site scales with ${\cal O}(D_W^2)$. The corresponding complexity, in the case of the quantity $\langle\langle\hat\rho|\mathcal{L}^{\dagger}\mathcal{L}|\hat\rho\rangle\rangle$, is sketched in Fig.~\ref{Graph}(e) and would naively scale as ${\cal O}(D_W^4)$. More specifically, the computational complexity according to Ref.~\onlinecite{MPSSch} (see Eq. (197)) is ${\cal O}(2d^2 D^3 D_W + d^4 D^2 D_W^2)$ for the $\mathcal{L}$ algorithm and would be ${\cal O}(2d^2 D^3 D_W^2 + d^4 D^2 D_W^4)$ for the $\mathcal{L}^{\dagger}\mathcal{L}$ case if one opted for directly using a MPO representation of the squared Liouvillian of bond dimension $D_W^2$. However, rather then constructing the $\mathcal{L}^{\dagger}\mathcal{L}$ MPO, one may improve the second approach by storing only $\mathcal{L}$ and carrying out the matrix multiplication by $\mathcal{L}^{\dagger}$ ``on the fly'' at each optimization step. This would reduce the complexity of the $\mathcal{L}^{\dagger}\mathcal{L}$ approach to
${\cal O}(2(d^2 D^3 D_W^2 + d^4 D^2 D_W^3)$, which is still however one $D_W$-factor slower than the $\mathcal{L}$ approach. For models such as bilinear-biquadratic hamiltonians, or even XYZ models with slightly involved dissipative processes, the MPO representation of $\mathcal{L}$ can reach bond dimension easily exceeding $D_W\approx10$. The present strategy may thus easily lead to a computational gain of more than one order of magnitude. Furthermore, in most systems of interest, $\mathcal{L}$ is a very sparse matrix, and this computational advantage is partly spoiled when instead adopting the generally less sparse squared Liouvillian. Finally, the bond dimension of the Liouvillian MPO has a relevant computational impact also on the iterative solution of the on-site eigenvalue problem at each site of the chain, in cases where the matrix is not fully stored and the linear operator is instead applied to vectors in a functional fashion. In these cases, for an MPO with bond dimension $D_W$, the complexity associated to the matrix-to-vector multiplication is ${\cal O}(D^3D_Wd^2+D^2D_W^2d^2+D^3D_W^2d^4)$ (see equation (201) of~\cite{MPSSch}), again highlighting the importance of using an MPO with minimal bond dimension. These considerations led us to explore the possibility of finding the NESS by directly searching for the null eigenvalue of $\mathcal{L}$.}

In the MPS-DMRG algorithm, all matrices $A$ and $B$ in (\ref{RhoMixed}) are kept constant, and the inequality (\ref{Func}) can then be cast into an on-site linear problem for the optimization of $\varrho$. To this purpose, we introduce the on-site Liouvillian operator $\mathcal{L}_l$ for site $l$, which is a rank-six tensor obtained from the quantities $\mathcal{L}$ and $|\hat\rho\rangle\rangle$ by contracting all indices associated to the other lattice sites, as depicted in Fig.~\ref{Graph}(c). The minimization of the norm functional (\ref{Func}) is then achieved by solving the local problem $\mathcal{L}_l\varrho=0$ successively on each site of the chain, sweeping along the chain in both directions until convergence to the null eigenvalue is reached. To this purpose, we solve the local problem by computing the complex eigenvalue of $\mathcal{L}_l$ closest to a small target scalar value. This scalar must be chosen much smaller than all energy scales characterizing the problem, in order to achieve convergence to the null eigenvalue of $\mathcal{L}$. Convergence is achieved after a sufficient number of sweeps, and by choosing bond dimensions large enough to accurately model the quantum correlations arising in the NESS. For the eigenvalue problem, we adopted here the Shift-and-Invert Arnoldi method, which is most efficient for small magnitude eigenvalues. The method has yielded in our tests the most stable and efficient realization of the algorithm. Due to the matrix inversion however, the Shift-and-Invert method requires the full storage of the local Liouvillian, i.e. a memory cost ${\cal O}(D^2d^2\times D^2d^2)$. {In cases where this memory cost cannot be afforded, it is still possible to adopt direct iterative schemes. The ARPACK library in particular~\cite{Softs} makes non-inverting versions of the Arnoldi method available. Our experience is that, while solving the storage problem, these methods are generally considerably slower and less stable -- though only marginally -- than the shift-and-invert method.}

In general, a matrix diagonalization targeting small complex eigenvalues is usually characterized by slow convergence. To overcome this limitation in the present case, and increase efficiency, we start the computation using a small bond dimension, and allow it to increase gradually along the sweeps, by each time padding the larger density matrix with zeros.
Lastly, we have found that the algorithm could become unstable when directly targeting very small dissipation rates (compared to the Hamiltonian energy scale). To ensure the stability of our implementation in such cases, we start the computation using larger dissipation rates, and let them decrease exponentially towards the desired values along the sweeps. In practice, in our tests we started the computation with values of $D$ between 5 and 10 and run several tens of sweeps, while gradually decreasing the dissipation rates if needed. We observed that this first phase can be sped up significantly by restricting the number of iterations of the Shift-and-Invert algorithm to less than 10. After this first phase has converged, we refine the result by allowing the bond dimension to increase gradually, while at the same time increasing the number of Shift-and-Invert iterations in each step to a few hundreds. This second phase typically requires less than 10 sweeps to achieve full convergence.

\begin{figure}
\includegraphics[width=1\linewidth]{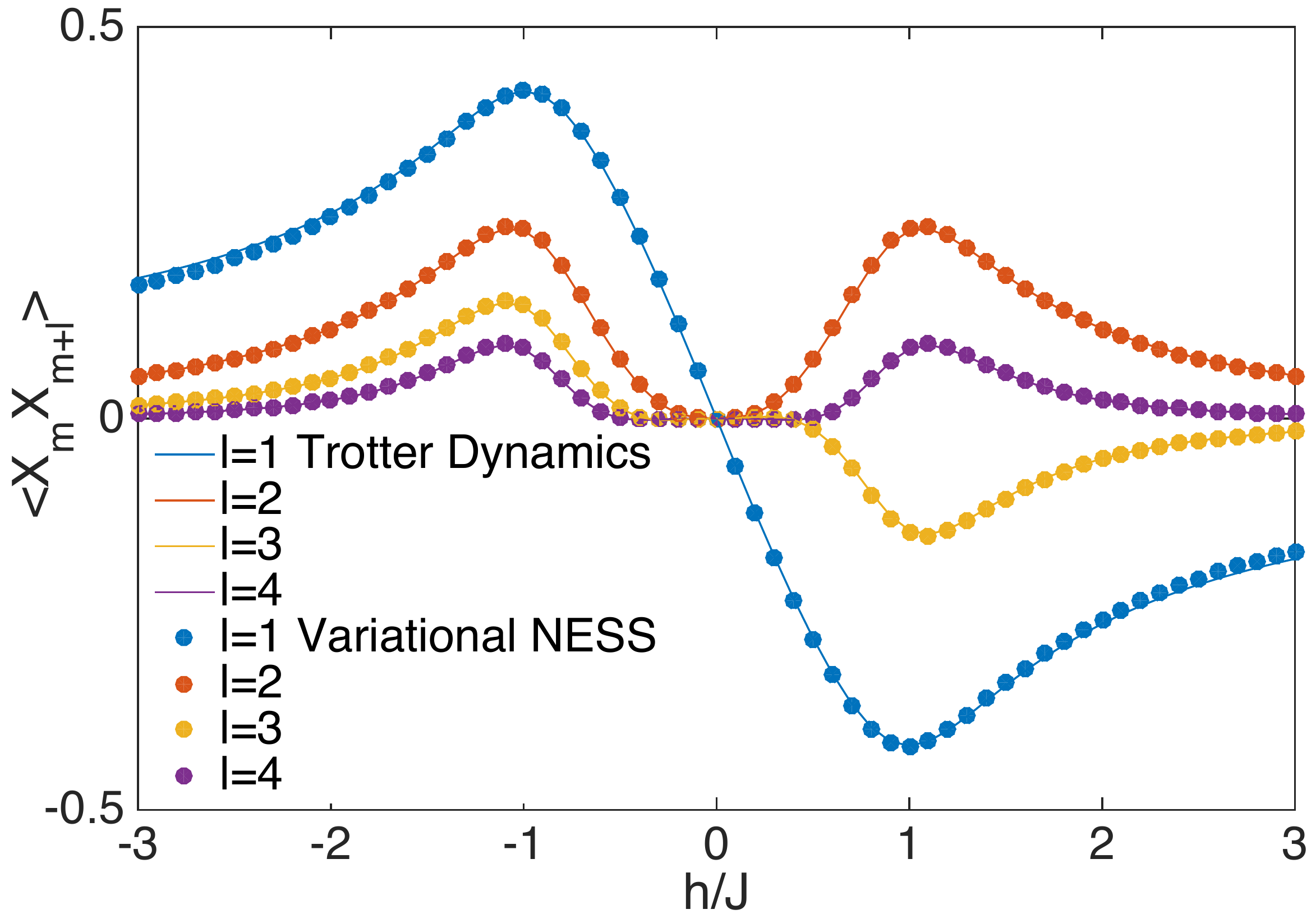}
\caption{Comparison between the spatial correlations $\langle \hat X_m \hat X_{m+l}\rangle$ for $l=1,\ldots,4$, computed by Trotter dynamics and through the direct variational MPS determination of the NESS. Parameters were set as $V=0$ and $\gamma=J$, and the array length was $N=15$. For the Trotter dynamics, the time step was set as $dt=0.1J$ and time integration was carried out until $T=10/\gamma$. For both the Trotter dynamics and the variational NESS methods the bond dimension was at most $D=20$. These results also displayed perfect agreement with MCWF calculations (not shown).}\label{TrotterVarNESS}
\end{figure}

Since the introduction of MPS modelling of mixed states, the issue of preserving the positivity of the density matrix has been discussed~\cite{MPOVer} and shown to be NP-hard to verify~\cite{EisertMPO}. It should be noted that only very recently a local purification scheme for the Trotter evolution has been proposed in~\cite{Simone} which guarantees positivity of the density matrix. However, we stress that for the cases we have considered we never encountered convergence to an MPS that presented unphysical results and we never had to reinforce the density matrix properties which, instead systematically result from the convergence of the algorithm.

Note also, that in the MPS approach the state is normalized according to the euclidean norm, i.e. $\langle\langle\hat\rho|\hat\rho\rangle\rangle=1$. Thus, in general, the condition on the trace $\mathrm{Tr}(\hat\rho)=\langle\langle \mathbb{I}|\hat\rho\rangle\rangle=1$ is not automatically fulfilled, and the expectation value of an arbitrary observable $\hat O$ must be evaluated as $\langle \hat O\rangle={\mathrm{Tr}(\hat \rho \hat O )}/{\mathrm{Tr}( \hat\rho )}={\langle\langle \mathbb{I}|\mathbb{I}\otimes \hat O|\hat\rho\rangle\rangle}/{\langle\langle \mathbb{I}|\hat\rho\rangle\rangle }$.

\section{Results}

As a test of the method, we simulate a driven-dissipative quantum Ising chain~\cite{Ima,Keeling}, described by the Hamiltonian
\begin{equation}
{\cal{\hat{H}}}=\sum_i\left[h \hat Z_i + J \hat X_{i} \hat X_{i+1} + V \hat X_{i} \hat X_{i+2}\right]\,,
\end{equation}
with $h$ being a local effective magnetic field, and $J$ and $V$ respectively the coupling between nearest neighbours and next nearest neighbours. The dissipative part is provided by transition operators $\hat K=\sqrt{\gamma}(\hat X-i\hat Y)/2$ at each site, with $\hat X$, $\hat Y$ and $\hat Z$ being Pauli matrices and $\gamma$ the dissipation strength.

\begin{figure}
\includegraphics[width=1\linewidth]{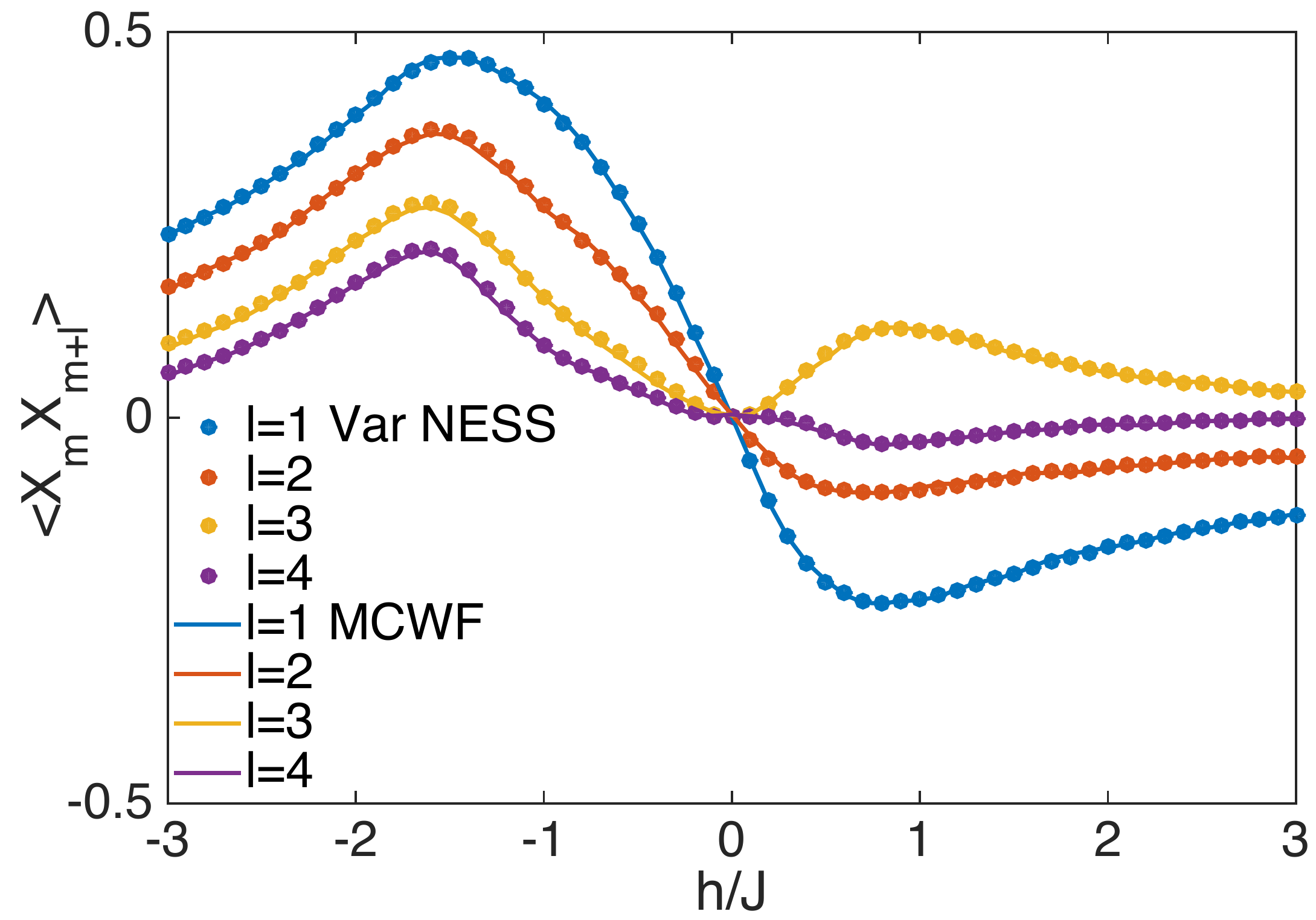}
\caption{Comparison between the spatial correlations $\langle \hat X_m \hat X_{m+l}\rangle$ for $l=1,\ldots,4$, computed by MCWF and through the direct variational MPS determination of the NESS. Parameters were set as $V=0.5J$ and $\gamma=J$, and the array length was $N=15$. The MCWF simulation was performed with 1000 trajectories and time integration was carried out until $T=10/\gamma$. For the variational NESS calculations, the bond dimension was at most $D=30$.}\label{MCWFVarNESSnnn}
\end{figure}

{Our study focuses on three paradigmatic cases and we initially assume a small system size (15 sites), to allow for a direct comparison with MCWF simulations. 
The MCWF unravels the master equation for density matrix into stochastic pure state trajectories in the Hilbert space. Dissipation is accounted for by non-hermitian terms in the Hamiltonian, while the corresponding fluctuations are enforced by random ``quantum jumps'' generated with a probability proportional to the square root of each dissipation rate. The method is described in detail in~\cite{MCWF1,MCWF2}, and we specifically adopted the QuTiP toolbox~\cite{QuTiP} for all MCWF calculations.
We also compare to the standard Trotter MPS evolution~\cite{Keeling} for benchmarking the method.}

In the first case, nearest neighbour couplings are considered, as in Ref.~\onlinecite{Keeling}. In Fig.~(\ref{TrotterVarNESS}), the results for the correlations $\langle \hat X_m \hat X_{m+l}\rangle$ for $l=1,\ldots,4$, obtained both using Trotter dynamics and the variational method are shown. They coincide perfectly with each other and with the data obtained in Ref.~\onlinecite{Keeling}. In particular, the system displays ferromagnetic order for negative external field and anti-ferromagnetic order for positive external field. The small discrepancy observed between the data obtained with the two methods is simply due to the Trotter error. For this case, $\gamma=J$ and the driven-dissipative time evolution is well handled by the Trotter dynamics, which is therefore the method of choice, as the time scale to reach the NESS is short and the resulting simulation turns out to be much faster than the variational method. This consideration holds in general, in cases with next-neighbour couplings and sufficiently fast dissipation rates.

\begin{figure}
\includegraphics[width=1\linewidth]{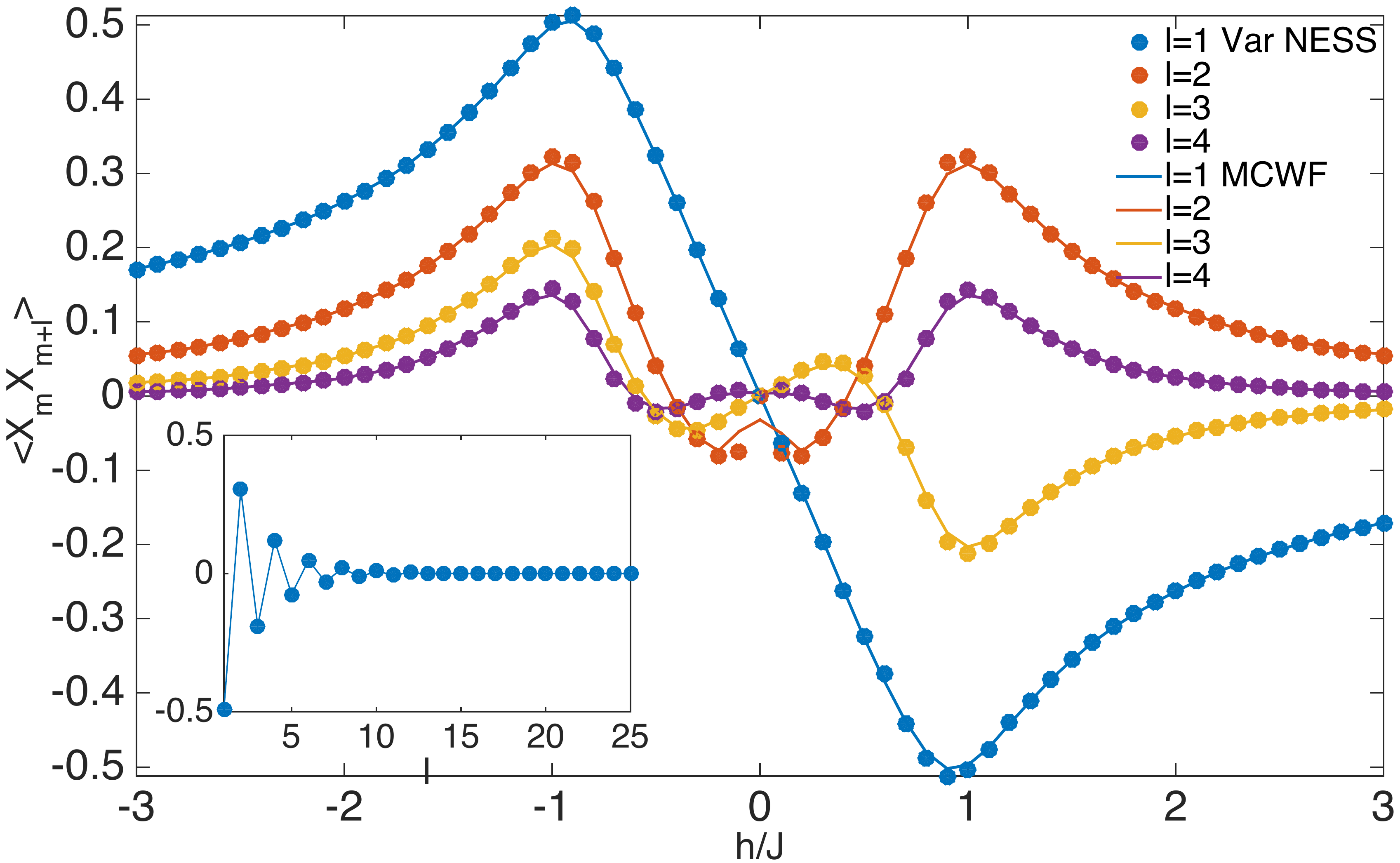}
\caption{Comparison between the spatial correlations $\langle \hat X_m \hat X_{m+l}\rangle$ for $l=1,\ldots,4$, computed by MCWF and through the direct variational MPS determination of the NESS, in the case of weak dissipation. Parameters were set as $\gamma=0.1J$, and the array length was $N=15$. The MCWF simulation was performed with 16 trajectories and time averages were taken from $t=10/\gamma$ and until $t=100/\gamma$. For the variational NESS the bond dimension was at most $D=50$. The inset shows the correlations $\langle \hat X_m \hat X_{m+l}\rangle$ as a function of $l$ with $h=J$ for a system of $N=50$ sites and bond dimension $D=60$. This system size lies beyond the computational reach of the MCWF method.}\label{MCWFVarNESSLowDiss}
\end{figure}

The second case we study, is that of a system with longer range couplings. In this case the usual Trotter dynamics cannot be employed and thus the variational NESS becomes the natural method of choice. In Fig.~\ref{MCWFVarNESSnnn} we compare results obtained with the MCWF and variational methods. Once again, we obtain a very good agreement between the two methods, even for small bond dimension. The next nearest neighbour coupling amplifies the ferromagnetic correlations, while having a sizeable effect on the anti-ferromagnetic side. By comparing Fig.~\ref{MCWFVarNESSnnn} and Fig.~\ref{TrotterVarNESS} we see that the next-nearest-neighbour correlation $\langle \hat X_m \hat X_{m+2}\rangle$ changes sign and we only observe anti-correlation at longer distance ($\langle \hat X_m \hat X_{m+3}\rangle$). We argue that, when adding genuinely long ranged couplings, the anti-ferromagnetic order in the positive external field sector might be completely suppressed.

As the third case, we simulate the same model in presence of a small dissipation rate. In this case, dynamical methods will become less effective and converge slowly. We have observed that the variational method in this case could become unstable. This issue was however completely removed by adopting a gradual decrease of the dissipation rate along the sweeps, as discussed previously.
In this case, the small dissipation rate results in increased correlations, both in the ferromagnetic and anti-ferromagnetic case, as show in Fig.~\ref{MCWFVarNESSLowDiss}. It is also interesting that nontrivial correlations emerge for very small external field showing that there are still novel regimes to be explored for these driven dissipative systems.
The inset in Fig.~\ref{MCWFVarNESSLowDiss} shows the correlations $\langle \hat X_m \hat X_{m+l}\rangle$ as a function of $l$, computed for a longer system with $N=50$ sites. The combination of a quasi-local hamiltonian with an on-site dissipation mechanism seems to generally lead to an exponential decay of the correlations. This setting typically holds for driven dissipative optical systems such as coupled optical cavities. This result suggests that the present method may efficiently model the NESS of long one-dimensional systems, already at moderate bond dimension.

 \section{Conclusion}

In conclusion, we have presented an efficient implementation of the variational principle for the NESS of one-dimensional driven-dissipative quantum systems using an MPO ansatz for the density matrix. The computational overhead of the method scales as a power law both in the dimension of the Hilbert space and in the bond dimension of the MPO. Vectorization allows to map the problem onto an effective linear eigenvalue problem, that can be then solved using a MPS-DMRG approach. We have applied the method to a model spin chain as a test, under various assumptions for the parameters. As compared to direct integration of the system dynamics, the present approach brings considerable advantage in cases where the dissipation rates are slow compared to the Hamiltonian energy scale. In particular, through a slow tuning of both the MPS bond dimension and the dissipation rates towards the target values, numerical stability and convergence to the physical NESS is achieved in all cases that we have studied. Also, the method gives access to systems with long-range couplings, for which the standard Trotter dynamics cannot be employed. In such cases, new algorithmic approaches to direct time integration have very recently emerged~\cite{Unify,Pollman}. The direct comparison between the present approach and these new developments is left as a venue for future investigations.

Modeling nonlinear driven-dissipative quantum systems generally represents a major challenge, as these systems combine the inherent difficulty in correctly describing quantum correlations to the nonequilibrium character of their approach to stationarity. This difficulty emerges, in particular, when dynamical critical phenomena and quantum phase transitions occur. Then, quantum correlations typically acquire a long spatial range and may even decay algebraically~\cite{StatesPhases}. Methods relying on the MPS ansatz are in these cases an ideal tool, as they provide control over the spatial range of quantum correlations through the bond dimension, while preserving a power-law computational complexity. In this framework, the method presented in this work holds promise as a powerful tool for the study of emergent quantum phenomena in nonequilibrium open quantum systems.

\acknowledgements
We acknowledge enlightening discussions with Fiona Seibold during the initial stage of this work. {We also aknowledge a very proficuous contact with Mari-Carmen Ba\~nuls who pointed out a more accurate estimate of the computational complexity}.

\end{document}